\documentclass[final,5p,times,twocolumn]{elsarticle}

\usepackage{amssymb}

\journal{Physics Letters B}

\usepackage{color}
\usepackage[normalem]{ulem}

\begin{document}

\begin{frontmatter}



\title{In-medium enhancement of the modified Urca neutrino reaction rates}

\author{Peter S. Shternin}

\address{Ioffe Institute, Politekhnicheskaya 26, 194021 St.~Petersburg, Russia}

\author{Marcello Baldo}
\address{Instituto Nazionale di Fisica Nucleare, Sez. di
Catania, Via S. Sofia 64, 95123 Catania, Italy}

\author{Pawel Haensel}
\address{N. Copernicus Astronomical Center, Polish Academy of
Sciences, Bartycka 18, PL-00-716 Warszawa, Poland}

\begin{abstract}
We calculate modified Urca neutrino emission rates
in the dense nuclear matter in neutron star cores. We find that these rates are strongly enhanced in the beta-stable matter in regions of the core close to the direct Urca process threshold. This enhancement can be tracked to the use of the in-medium nucleon spectrum in the virtual nucleon propagator.
We describe the in-medium nucleon scattering in the non-relativistic Bruckner-Hartree-Fock framework taking into account two-body as well as the effective three-body forces, although the proposed enhancement does not rely on a particular way of the nucleon interaction treatment. Finally we suggest a simple approximate expression for the emissivity of the neutron branch of the modified Urca process that can be used in the neutron stars cooling simulations with any nucleon equation of state of dense matter.
\end{abstract}


\end{frontmatter}


\section{Introduction}
\label{S:intro}
Neutron stars (NSs) are the densest stars in the Universe. Their cores contain
cold superdense matter with densities reaching several times the nuclear saturation density
 $n_0=0.16$~fm$^{-3}$.
 NSs occupy the unique location in the QCD
phase diagram, currently unreachable by the modern ground-based
experimental studies, and as such the composition and equation of state (EOS)
in NS interiors is largely unknown \cite{HPY2007Book}. On the other hand, NSs have extremely large diversity of astrophysical manifestations spanning the whole electromagnetic \cite{Degenaar2018arXiv}, and, recently,  gravitational wave spectrum \cite{Abbott2017ApJ,Abbott2017PhRvL}. It is believed that confronting the observational data with the results of theoretical modeling of various processes in NSs one can constrain uncertain properties of their interiors and, as a consequence, increase the knowledge  about the fundamental interactions in dense matter. That is why NS studies attract constant
attention.

One of a few insights into the
uncertain physics of the NS interiors comes from the study of the
thermal evolution of these objects, either isolated or in binary systems, see, e.g. \cite{YakovlevPethick2004ARA,Degenaar2018arXiv}. One of the main cooling regulators, alongside with the surface electromagnetic emission,  is the neutrino emission from the NS bulk. For sufficiently hot NSs the latter is in fact the main ingredient of the NS cooling theory \cite{YakovlevPethick2004ARA}.

There is a big diversity of neutrino generation processes
inside NSs, with the
liquid stellar core being the source of the strongest ones
\cite{Yakovlev2001physrep}. Operation of these processes and their rates inevitably depend on the NS EOS and composition. For instance, the most
powerful mechanism of the neutrino emission in nucleon cores of
NSs, the so-called direct Urca process, consists of a pair of charged weak current
reactions ${\rm n}\to {\rm p}+\ell+\bar{\nu}_\ell,\, {\rm
p}+\ell\to {\rm n}+\nu_\ell$, where $\ell$ is a lepton, electron
or muon, and $\nu_\ell$ is the corresponding neutrino. Strong degeneracy of the NS matter puts a fundamental restriction on the direct Urca process requiring that $p_{\mathrm{F}n}<p_{\mathrm{F}p}+p_{\mathrm{F}\ell}$, where $p_{\mathrm{F}i}$ is the Fermi momentum of the $i$ species. This suggests that the proton fraction should be sufficiently high for the direct Urca process to operate. Therefore, the direct Urca process can proceed in sufficiently heavy NSs with central density larger than some threshold density $n_\mathrm{dU}$.  Different EOSs  predict different  $n_\mathrm{dU}$. In lighter NSs and for some EOSs in all NSs up to maximally massive ones, direct Urca processes are forbidden and other reactions come into a play. In this case, the basic neutrino
emission mechanisms in the (non-superfluid) NSs
involve nucleon collisions. The strongest
process of this type is the modified Urca process which also proceeds via the charged weak current and is given by a
pair of reactions:
\begin{equation}
{\rm n}+{\rm N}\to {\rm p}+{\rm N}+\ell+\bar{\nu}_\ell,\quad
  {\rm p}+{\rm N}+\ell\to {\rm n}+{\rm
  N}+\nu_\ell.\label{eq:murca}
\end{equation}
Here N is the additional nucleon
which relaxes the momenta restrictions. The companion nucleon bremsstrahlung processes which involve neutral weak currents are order of magnitude less powerful \cite{Yakovlev2001physrep}.

The standard benchmark for the treatment of modified Urca reactions in NS physics is the work by Friman and Maxwell \cite{FrimanMaxwell1979ApJ}. It is based on the free one-pion exchange (OPE) model and the central part (nuclear correlations) is described by a certain set of Landau parameters. Several studies expand results of Ref.~\cite{FrimanMaxwell1979ApJ}, basically focusing on the improvement of the in-medium effects treatment, see, e.g., Ref.~\cite{Schmitt17}. In particular, Ref.~\cite{Blaschke1995MNRAS} estimated the effect of the replacement of the free one-pion exchange interaction with the in-medium $T$-matrix and found some reduction of the emissivity. Recent study \cite{Niri2016PhRvC} employed independent pair approximation extending Ref.~\cite{Haensel1987A&A}. They calculated pair correlation function in the variational approach accounting for two-body as well as three-body forces. The final result turned out to be not so far from those of Ref.~\cite{FrimanMaxwell1979ApJ}.

In a series of papers staring from Ref.~\cite{Voskresensky1986JETP} the scenario named `medium modified Urca' was developed, see Refs.~\cite{Voskresensky2001LNP, Migdal1990PhR} for review. The basis of this scenario is the pion-exchange model of the interaction with a strong softening of the in-medium pion (medium-modified one-pion exchange, or MOPE). This leads to the strong enhancement of the modified Urca rates. Moreover, it was argued that the strongest subprocess involving  charged weak current is the conversion of the virtual charged pion to virtual neutral pion with the emission of the lepton pair. However these results are strongly dependent on a particular choice of the model paramaters. In
some cases a strong softening of the pion mode is the precursor of the real pion condensation at higher density \cite{Voskresensky2001LNP}.

In this letter we show that all previous studies missed an important piece of a picture. Specifically, we argue that the account of the nucleon potential energy in the medium amplifies considerably the modified Urca rates. The proposed amplification is universal, resulting only from the requirement of the beta-equilibrium, and the importance of this amplification increases when the density gradually approaches $n_\mathrm{dU}$.

The paper is organized as follows. In Sec.~\ref{S:formalism} we briefly present the standard formalism for the calculations of the modified Urca rates. The in-medium nucleon propagator is discussed in Sec.~\ref{S:prop}; this section contains the main result of our work. In Sec.~\ref{S:Gmat} we outline the adopted model for in-medium scattering. We describe the nucleon interaction by means of the $G$-matrix of the Brueckner-Hartree-Fock theory constructed on top of the realistic nucleon potential with inclusion of the effective three-body forces. We  discuss our results and illustrate  their effect on the model cooling calculations in Sec.~\ref{S:discuss} and conclude in Sec.~\ref{S:conclusion}.

For concreteness, we focus on the neutron branch of the modified Urca process, where  $N=n$ in Eq.~(\ref{eq:murca}), although the obtained results are qualitatively applicable to the proton branch ($N=p$) as well. Effects of superfluidity and magnetic fields are not considered. Unless otherwise is indicated, we use
the natural unit system with $k_B=\hbar=c=1$.

\section{Formalism}
\label{S:formalism}
In the conditions appropriate for the NS cores below the direct Urca threshold, it is enough to describe nucleons in the non-relativistic quasi-particle approximation. The modified Urca emission rate can then be found from the Fermi golden rule for each of the reactions~(\ref{eq:murca}). The
detailed derivation is given, for instance, in the review
~\cite{Yakovlev2001physrep}. Under the conditions of beta-equilibrium,
the rates of forward and reverse reactions are equal, so one can
consider one reaction of the pair [we focus on the first reaction in (\ref{eq:murca})] and double the result. The neutrino emissivity is then
\begin{eqnarray}
  Q^{(\ell)}_\mathrm{MU}&=& 2\int \prod\limits_{j=1}^4 \frac{{\rm d}
  \mathbf{p}_j}{(2\pi)^3}\int \frac{{\rm d} \mathbf{p}_\ell}{(2\pi)^3}
  \int \frac{{\rm d} \mathbf{p}_\nu}{(2\pi)^3}\; \varepsilon_\nu
  (2\pi)^4\nonumber\\
  &&\times \delta(E_f-E_i)\delta(\mathbf{P}_f-\mathbf{P}_i){\cal F} s {\cal
  M}_{fi},\label{eq:golden}
\end{eqnarray}
where $j$ enumerate nucleons, ${\cal F}=f_1 f_2(1-f_3)(1-f_4) (1-f_\ell)$ is the Pauli blocking factor with
$f_i=\left[1+{\rm
exp}\left((\varepsilon_i-\mu_i)/T\right)\right]^{-1}$ being the
Fermi-Dirac function, $\varepsilon_i$, $\mathbf{p}_i$, and $\mu_i$ are the quasiparticle energy, momentum, and chemical potential, respectively,  $T$ is the temperature,  ${\cal M}_{fi}\equiv\sum_{\rm spins}
|M_{fi}|^2$ is the squared matrix element of the process, summed
over the spin states, $s=2^{-1}$ is the symmetry factor that accounts for the double-counting of the same collision events,  $(E_f,\mathbf{P}_f)$  and $(E_i,\mathbf{P}_i)$ are the total energy and momentum of the final and initial particles, respectively.

Since neutron star matter is strongly degenerate, all fermions except neutrinos in Eq.~(\ref{eq:golden}) are placed at the respective Fermi surfaces. The neutrino momentum is small (of the order of $T$), therefore it can be neglected in the momentum conservation and in the matrix element. These facts allow to decompose energy and angular integrations in Eq.~(\ref{eq:golden}) which greatly simplifies the calculations. As a result, the phase space averaging of the matrix element contains only four non-trivial angular integrations, see, e.g.,  \cite{Yakovlev2001physrep,Schmitt17,Kaminker2016Ap&SS}.

\begin{figure}[th]
\begin{center}
{\includegraphics[width=0.45\columnwidth,clip]{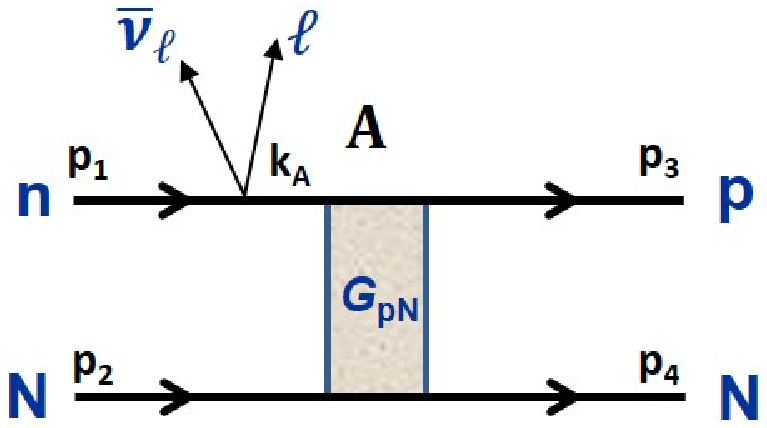}}
    {\includegraphics[width=0.45\columnwidth,clip]{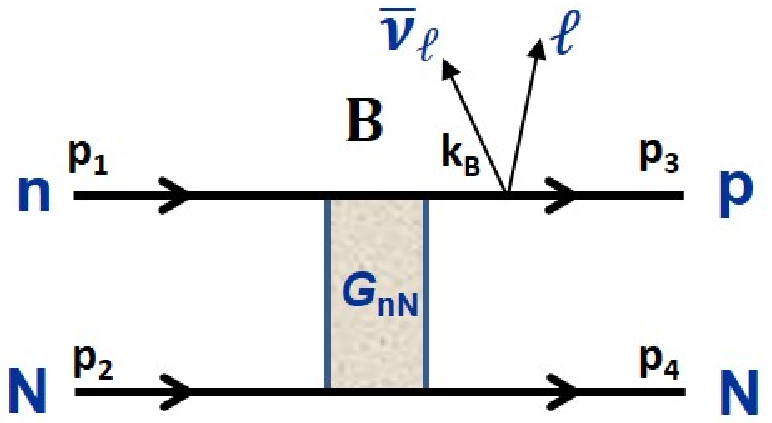}}
    \caption{ Direct external leg Feynmann diagrams A (left) and B (right) contributing to the
   modified Urca process. }
   \label{fig:nn}
\end{center}
\end{figure}

In the non-relativistic $V-A$ approximation, the weak interaction Lagrangian is
\begin{equation}\label{eq:WeakL}
{\cal L}=\frac{G_F\cos\theta_C}{\sqrt{2}} l^\mu \Psi_p^\dag\left(g_V\delta_{\mu,0}-g_A \delta_{\mu,i}\sigma_i\right)\Psi_n,
\end{equation}
where $G_F=1.17\times10^{-5}$~GeV$^{-2}$ is the Fermi coupling constant, $\cos\theta_C=0.975$ is the cosine of the Cabibbo angle, $\Psi_p$ and $\Psi_n$ are the proton and neutron spinors, respectively, $\sigma_{i}$, $i=1\dots 3$, are the Pauli matrices,
$g_V=1$ and $g_A\approx 1.26$ are nucleon weak vector and axial vector coupling constants, respectively. Lepton charged current is
$l_\mu=\bar{l}\gamma_\mu(1-\gamma_5)\nu$, where $l$ and $\nu$ are lepton and antineutrino Dirac spinors, $\gamma_\mu$ and $\gamma_5$ are Dirac matrices. Here we neglect additional contributions from the weak magnetism or induced pseudoscalar interactions \cite{Timmermans2002PhRvC}.

The basic direct diagrams which contribute to the modified  Urca
processes are given in Fig.~\ref{fig:nn} where the hatched blocks
represent the nucleon interaction. The amplitude corresponding to diagrams in Fig.~\ref{fig:nn}
is (neglecting neutrino momentum)
\begin{eqnarray}\label{eq:Mdir}
M_{fi}^{\mathrm{dir}}&=&l^\nu\left(\hat{\Gamma}_\nu {\cal G}_p(\varepsilon_1-\varepsilon_\ell,\mathbf{p}_1-\mathbf{p}_\ell) G_{pn}\right.\\
&&\left.+ G_{nn} {\cal G}_n(\varepsilon_3+\varepsilon_\ell,\mathbf{p_3}+\mathbf{p}_\ell)\hat{\Gamma}_\nu\right),\nonumber
\end{eqnarray}
where $\hat{\Gamma}_\nu$ is the weak vertex, which follows from (\ref{eq:WeakL}), ${\cal G}_{n(p)}$ is the neutron (proton) propagator, and $G_{pn}$ and $G_{nn}$ are the scattering matrices corresponding to the proton-neutron interaction for the diagram A and neutron-neutron interaction for the diagram B, respectively. The exchange diagrams correspond to the interchange of initial states $\{1\leftrightarrow 2\}$. In addition to the external-leg emission diagrams shown in Fig.~\ref{fig:nn}, there are other diagrams that can contribute to the modified Urca process, see, e.g., Refs.~\cite{Blaschke1995MNRAS,Hanhart2001PhLB,Schmitt17}. These diagrams correspond to intermediate state processes and generally are of the next order in the emitted pair momentum.  We do not consider these diagrams now and discuss them later in Sec.~\ref{S:discuss}.

The medium  enters Eq.~(\ref{eq:Mdir}) via the mean field in the nucleon propagators and the in-medium scattering matrices. Below we consider these effects separately.

\subsection{Nucleon propagator} \label{S:prop}
The quasi-particle propagator associated with the intermediate line in Fig.~\ref{fig:nn} is
\begin{equation}
  {\cal G}_N(E,k)=\left(E-\varepsilon_N(k)\right)^{-1},\label{eq:prop}
\end{equation}
where $\varepsilon_N(k)$ is the nucleon in-medium spectrum. Traditionally, after the work by Friman and Maxwell \cite{FrimanMaxwell1979ApJ}, one approximates ${\cal G}_N(E,k)\approx \pm\mu_\ell^{-1}$ where the sign depends on whether the emission occurs before scattering (diagram A) or after scattering (diagram B). However, this approximation misses the softening of the intermediate nucleon line that occurs in the beta-equilibrium nuclear matter. Consider, for instance, the A diagram, where the intermediate line corresponds to the virtual proton. Neglecting neutrino energy and placing all quasiparticles in external lines on the respective Fermi surfaces, one finds for the A diagram $E_\mathrm{A}=\mu_n-\mu_\ell=\mu_p$, where the second equality is due to the beta-equilibrium condition. At the densities above the direct Urca threshold, the intermediate line contains the pole in the allowed phasespace which opens the direct Urca process. Momentum conservation implies $\mathbf{k}_\mathrm{A}=\mathbf{p}_1-\mathbf{p}_\ell$, hence $p_{\mathrm{F}n}-p_{\mathrm{F}\ell}\leq k_\mathrm{A} \leq p_{\mathrm{F}n}+p_{\mathrm{F}\ell}$. Therefore, below the direct Urca threshold, $k_\mathrm{A}$ is always above the proton Fermi surface $p_{\mathrm{F}p}$. In case of the backwards emission, however, $k_\mathrm{A}$ can be sufficiently close to $p_{\mathrm{F}p}$ or, in other words, to the pole of the propagator, $\varepsilon_p(k)=\mu_p$, leading to the enhancement of the emission amplitude in that part of the phasespace. Similar arguments hold for the B diagram, where the intermediate nucleon is a neutron and it is propagating below the Fermi surface. We stress, that the enhancement arises from the proper use of the single nucleon spectrum, in whatever approximation it is calculated provided the consistency with beta equilibrium is kept. This results in a large difference between potential energies of the neutron and proton quasiparticles in NS matter which can not be neglected.

In order to quantify the enhancement that comes from the use of the beta-equilibrium propagator instead of the usual approximation, it is instructive to consider the factors
\begin{equation}
  R_{\rm dir}=\left\langle\left|\frac{\mu_\ell}{\mu_{\rm N}-\varepsilon_{\rm
  N}(k)}\right|^2\right\rangle,
\end{equation}
where the angular brackets denote the phasespace averaging. These factors will describe the modification of the emission corresponding to the direct diagram A (with $N=p$) or B (with $N=n$) if the scattering amplitude weakly depends on the angular variables. In order to calculate these factors we need the quasiparticle
spectrum. We adopt the following non-relativistic expression
\begin{equation}
  \varepsilon_N(k)-\mu_N=\frac{k^2}{2 m_N^*}-\frac{p_{\mathrm{F}N}^2}{2m_N^{*}},
\end{equation}
where $m_N^*$ is the nucleon effective mass at the Fermi surface. This approximation may not be very accurate far off-shell, however as discussed, the main contribution to the rate is expected near the Fermi surface, where this form is sufficient. The phasespace averaging for the direct diagrams reduces to the integration over the allowed absolute values of intermediate momentum \cite{Kaminker2016Ap&SS} leading to
\begin{eqnarray}
    R_{\rm dir}^{(A)}&=&
    \frac{2 m_p^{*2} \mu_\ell^2}{p_{\mathrm{F}\ell}}\int\limits_{p_{\mathrm{F}n}-p_{\mathrm{F}\ell}}^{p_{\mathrm{F}n}+p_{\mathrm{F}\ell}}
    \frac{{\rm d} k_\mathrm{A}}{(p_{\mathrm{F}p}^2-k_\mathrm{A}^2)^2},\label{eq:RdirA}\\
    R_{\rm dir}^{(B)}&=&
    \frac{2 m_n^{*2} \mu_\ell^2}{p_{\mathrm{F}\ell}p_{\mathrm{F}p}}\int\limits_{p_{\mathrm{F}p}-p_{\mathrm{F}\ell}}^{p_{\mathrm{F}p}+p_{\mathrm{F}\ell}}
    \frac{k_\mathrm{B}\, {\rm d} k_\mathrm{B}}{(p^2_{\mathrm{F}n}-k_\mathrm{B}^2)^2}.\label{eq:RdirB}
\end{eqnarray}
These integrals can be easily taken analytically; we do not give explicit expressions here for brevity. Both $R$-factors have the asymptotic behavior $R_\mathrm{dir} \propto (p_{\mathrm{F}n}-p_{\mathrm{F}p}-p_{\mathrm{F}\ell})^{-1}$ in the vicinity of the direct Urca threshold. This result shows that there is a general density dependence of the modified Urca rate that leads to significant enhancement near the direct Urca threshold. In addition to the direct contributions from the diagrams A and B, the final expression contains also the interference term between these diagrams and also the interference term with the exchange contribution. In these terms, the intermediate nucleon
softening occurs in different parts of the phasespace for two interfering amplitudes. As a consequence, the enhancement of the interference terms is weaker, being of the logarithmic order in momenta difference. Calculations show that the interference terms can be neglected in practice, see below.

\subsection{Nucleon interaction} \label{S:Gmat}
We choose to treat the in-medium nucleon scattering in the framework of the non-relativistic Brueckner-Hartree-Fock (BHF) approach \cite{Baldo1999Book}. The in-medium scattering matrix, or $G$-matrix, is found from the solution of the Bethe-Goldstone equation.
In what follows we omit nucleon isospin indices for the sake of simplicity, and refer to Sec.~3.2 of Ch.~8 of Ref.~\cite{Baldo1999Book} for a complete
form of the relevant equations. Bethe-Goldstone equation reads
\begin{equation}\label{eq:BBG}
G[n_\mathrm{B};\omega]=V+\sum\limits_{k_a,k_b} V \frac{|k_a k_b\rangle{\cal Q}\langle k_a k_b|}{\omega-e(k_a)-e(k_b)} G[n_\mathrm{B};\omega],
\end{equation}
where $V$ is the bare nucleon interaction, $\omega$ is the starting energy, ${\cal Q}$ is the Pauli operator. Single-particle energy $e(k)$ in turn depends on the $G$-matrix\footnote{Notice, that the true quasiparticle spectrum $\varepsilon_N(k)$ in Eq.~(\ref{eq:prop}) in general differs from the BHF lowest-order single-particle potential $e(k)$ \cite{Baldo1999Book}. }:
\begin{equation}\label{eq:SPP}
e(k)=\frac{k^2}{2 m_N} + \mathrm{Re}\sum\limits_{k'\leq p_\mathrm{F}} \langle kk'|G[n_\mathrm{B}; e(k)+e(k')]|kk'\rangle_a,
\end{equation}
where subscript $a$ means antisymmetrization of the wavefunction, therefore  Bethe-Goldstone equation needs to be solved iteratively. In Eqs.~(\ref{eq:BBG})--(\ref{eq:SPP}) the so-called continuous choice of the single-particle potential is adopted \cite{Baldo1999Book}.

For the bare nucleon interaction on the two-body level we use Argonne $v18$ potential \cite{Wiringa1995PhRvC}. We also include three-body forces which are necessary in the non-relativistic theory to reproduce the empirical saturation point of the symmetric nuclear matter. Three body forces are included in the effective way to the two-body interaction via averaging over the third particle \cite{Grange1989}. Specifically, here we use the phenomenological Urbana IX model  \cite{Carlson1983NuPhA} adjusted to give the correct saturation point with the $v18$ potential \cite{Baldo2008PhLB}.
Bethe-Goldstone equation was solved in a partial wave representation with inclusion of the total nucleon pair angular momenta up to $J=12$. In Eq.~(\ref{eq:Mdir}) half-on-shell $G$-matrix is needed, where the starting energy is equal to the energy of one -- incoming or outcoming -- pair of the nucleons, but not to another.

\section{Results and discussion}
\label{S:discuss}
 \begin{figure}[th]
 \begin{center}
 \includegraphics[width=0.8\columnwidth]{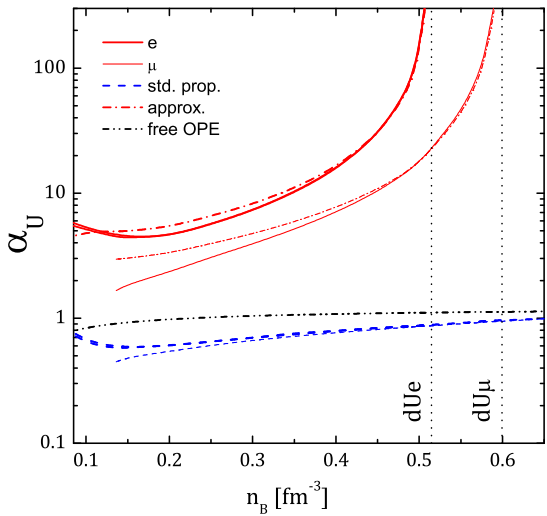}\vskip0.4cm
    \caption{Dimensionless rate $\alpha_{\rm U}$ for the modified Urca processes with electrons (thick lines) and muons (thin lines). Dashed lines show the results calculated using the in-medium $G$-matrix for the nucleon scattering, but the standard expression for the nucleon propagator. Dash-dotted lines show the simple approximation (\ref{eq:alpha_appr}). Double-dot-dashed line represents the result in the free OPE model. Vertical dotted lines show the direct Urca threshold densities for electrons and muons. See text for details.}
    \label{fig:alpha_e}
    \end{center}
 \end{figure}
It is convenient \cite{Yakovlev2001physrep},  after Friman and Maxwell \cite{FrimanMaxwell1979ApJ}, to normalize emissivity as
\begin{eqnarray}
  Q^{(\ell)}_{21}&=&8.1\, \frac{m_n^{*3} m_p^*}{m_u^4}
  \left(\frac{p_{\mathrm{F}\ell}}{\mu_\ell}\right)\left(\frac{n_p}{n_0}\right)^{1/3}
  T_9^8\, \alpha_U^{(\ell)},
  \label{eq:Qn}
\end{eqnarray}
where $Q^{(\ell)}_{21}\equiv Q^{(\ell)}_\mathrm{MU}/\left(10^{21}~{\rm erg}~{\rm s}^{-1}~{\rm cm}^{-3}\right)$, $T_9\equiv T/\left(10^9~\mathrm{K}\right)$, $m_u$ is the nuclear mass  unit, and dimensionless amplitudes $\alpha_U^{(\ell)}$ are proportional to
phasespace-averaged squared matrix element of the process,
$\langle {\cal M}_{fi}\rangle$. The normalization is governed by the free OPE interaction model adopted in Ref.~\cite{FrimanMaxwell1979ApJ} and, specifically, is
\begin{equation}
\langle {\cal M}_{fi}\rangle=64 G_F^2\cos^2\theta_C\left(\frac{f_\pi}{m_\pi}\right)^4 \frac{g_A^2}{\mu_\ell^2}\alpha^{(\ell)}_U,
\end{equation}
where $f^2_\pi\approx 0.08$ is the pion-nucleon coupling constant,  $m_\pi=139.6$~MeV is the charged pion mass. Calculations of the dimensionless amplitudes $\alpha_U^{(e)}$ (for mUrca process with electrons) and $\alpha_U^{(\mu)}$ (for mUrca process with muons) are shown in Fig.~\ref{fig:alpha_e} by thick and thin lines, respectively. Double-dot dashed line shows simplified free OPE result $\alpha_U^\mathrm{OPE}= 21 p_{\mathrm{F}n}^4/[16 (p_{\mathrm{F}n}^2+m_\pi^2)^2]\approx 1$, which is the same for electron and muon processes \cite{Yakovlev2001physrep}.  Dashed lines show the results of the present calculations where the $G$-matrix is used for the interaction, but the traditional nucleon propagator is used. Notice the small difference between the electron and muon lines as well. Solid lines include both the $G$-matirix and the nucleon propagator as described in Sec.~\ref{S:prop}. For the latter, we use the effective masses calculated from the BHF single-particle potential \cite{Baldo2014PhRvC}. One observes significant enhancement of the emissivity near the direct Urca thresholds, in agreement with the discussion in Sec.~\ref{S:prop}. In the present model, direct Urca threshold for electrons occurs at baryon density $n_\mathrm{B}=n_{\mathrm{dU}e}=0.515$~fm$^{-3}$ that correspond to a proton fraction of $x_p=0.137$ and at slightly higher density for muons, $n_{\mathrm{dU}\mu}=0.598$~fm$^{-3}$, $x_p=0.158$ \cite{Sharma2015A&A}.  Moreover, the emissivity is enhanced by a factor of few for all densities, even far from the threshold (compare solid and dashed lines). This is due to a large numerical values of the prefactors in Eqs.~(\ref{eq:RdirA})--(\ref{eq:RdirB}). The curves for muons in Fig.~\ref{fig:alpha_e} start from $n_\mathrm{B}=0.136$~fm$^{-3}$, the density of the muon appearance in our model.

\begin{figure*}[t]
\begin{center}
    {\includegraphics[width=0.4\textwidth,clip]{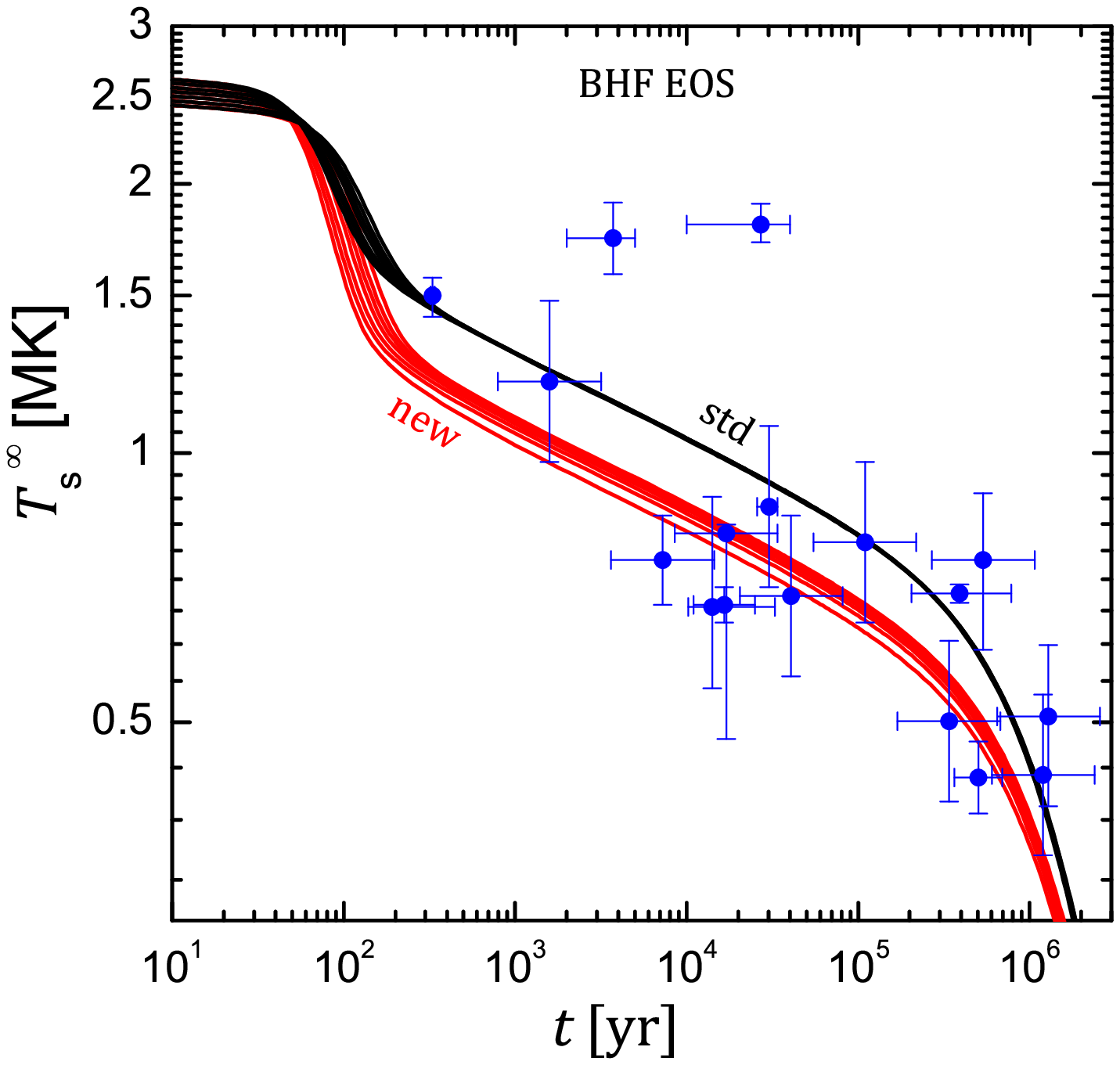}}
    \hspace{0.05\textwidth}
    {\includegraphics[width=0.4\textwidth,clip]{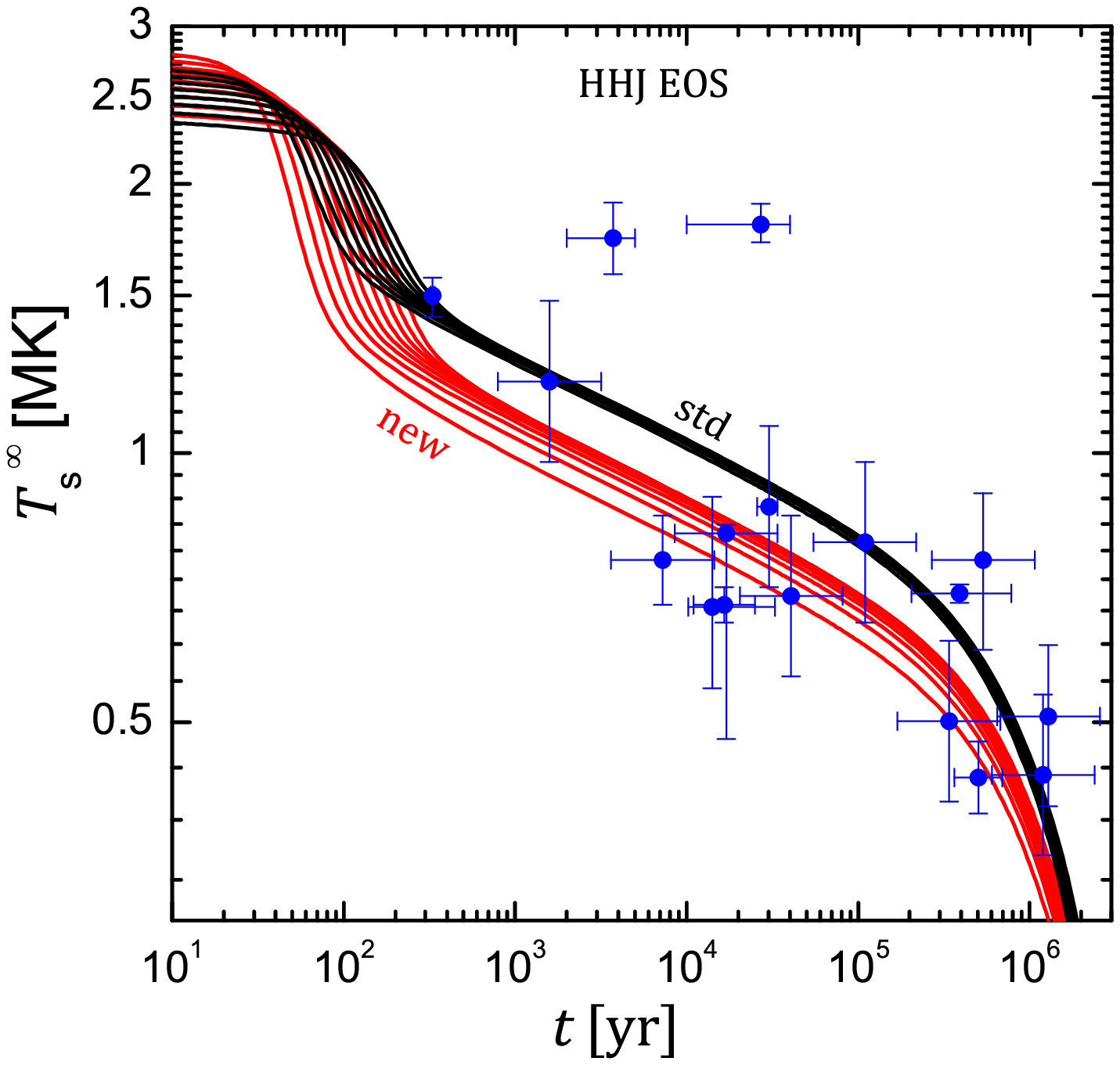}}
   \caption{Cooling curves (surface temperature $T_s^\infty$ vs. age $t$) of the isolated neutron stars. Groups of curves marked `std' show the results of the standard cooling simulations. Curves `new' show the results based on the rates from the present study. Blue error-crosses show observational data for several cooling NSs. Left panel corresponds to NS models with BHF EOS, the cooling curves span mass range $1.0-1.3\; M_\odot$ stepped by $0.05 M_\odot$. Right panel is for the HHJ EOS. Mass range for the curves is $1.0-1.8\; M_\odot$ stepped by $0.1M_\odot$. }
   \label{fig:cool}
\end{center}
\end{figure*}

The scattering amplitudes needed for the modified Urca calculations are half on-shell. However, the divergent structure of the propagator selects those parts of the phasespace where the closest to on-shell $G$-matrix is most important. Hence, close to the threshold, modified Urca rates are regulated by the non-radiative scattering rates. In this respect, our results resemble the soft electroweak bremsstrahlung theorem extensively discussed in the context of the nucleon bremsstrahlung neutrino emission in Refs.~\cite{Hanhart2001PhLB,Timmermans2002PhRvC}, see also the reviews \cite{Sedrakian2007PrPNP,Schmitt17} and references therein.
Soft electroweak bremsstrahlung theorem (SBT) states in analogy with the soft electromagnetic bremsstrahlung theorem that the emission rate in the leading (determined by the diagrams in Fig.~\ref{fig:nn}) and the first subleading order relates to the non-radiative scattering amplitudes. This theorem holds in the soft regime, where the radiated energy is small, therefore it is not strictly applicable to the modified Urca rates, where the radiated energy is basically the lepton energy $\mu_\ell$. Therefore SBT for the modified Urca processes is questionable \cite{Hanhart2001PhLB,Timmermans2002PhRvC,VanDalen2003PhRvC}. The present results point that at least in the leading order, the relations analogous to the SBT hold for the modified Urca rates close to the direct Urca threshold. It would be interesting to reveal to what extent this analogy holds.

Beyond SBT regime, next-order diagrams with respect to those in Fig.~\ref{fig:nn} can be important. These are, for instance,  rescattering diagrams where the lepton pair emission occurs in the intermediate nucleon line between two scatterings and the emission from the two-body axial vector current \cite{Hanhart2001PhLB,Migdal1990PhR}. The rescattering diagrams effectively are included in  Refs.~\cite{Niri2016PhRvC,Haensel1987A&A}
via the pair correlation functions. The results of these authors are not dramatically far from the standard results of Ref.~\cite{FrimanMaxwell1979ApJ} although the isolation of the rescattering contribution in their approach is not straightforward.

In the medium-modified Urca approach reviewed, e.g., in Ref.~\cite{Voskresensky2001LNP}, the third class of the diagrams where the emission comes from the intermediate pion line was found to dominate over the `external leg' emission described by the diagrams analogous to Fig.~\ref{fig:nn} (with the in-medium pion exchange line in place of the scattering matrix). The calculations of the external leg diagrams in their approach were also based on the standard nucleon propagators. Therefore the inclusion of the results presented here may change this conclusion.  We note, that the general MOPE enhancement would survive, since it is based on the softening of the intermediate pion which amplifies the external leg emission as well, as seen for the MOPE bremsstrahlung rates. Since, as discussed above, reinstall of the SBT can be expected in  certain sense, it is not clear if the dominance of the intermediate pion emission would survive.  Furthermore, the softening of the pion in the medium is at least questionable \cite{BrownRho2004}. Modification of the MOPE rates with account for the effects described in the present paper deserves future studies.

Dong et al. \cite{Dong2016ApJ} discussed the modification (suppression) of the modified Urca rates by the quasiparticle Fermi surface depletion quantified by the quasiparticle strength $z$. Modified Urca rates are proportional to the fourth power of $z$ leading to a certain reduction of the rates. At our level of the BHF theory, there is no depletion, i.e. $z=1$. Inclusion of the rearrangement contributions leads to $z<1$, however in this case, the nuclear scattering
needs  to be modified consistently. Moreover, the Fermi surface depletion is counter-balanced in a certain way by the corresponding increase in the nucleon effective masses  \cite{Schwenk2003NuPhA,Schwenk2004PhLB}. Therefore, in the present study we do not account for the $z$-factors and accordingly use the effective masses of the lowest order  Brueckner theory.

Returning to the present results, the modest density dependence of the dashed curves in Fig.~\ref{fig:alpha_e} allows to construct a simplistic approximate practical expression for the mUrca rate
\begin{equation}\label{eq:alpha_appr}
\alpha_U^\mathrm{approx} = 0.6 R_\mathrm{dir}^{(A)} + 0.2 R_\mathrm{dir}^{(B)}
\end{equation}
shown with dash-dotted lines in Fig.~\ref{fig:nn}. The difference between the results of the full calculations and the approximate expression does not exceed 20 percent at $n_\mathrm{B}>0.1$~fm$^{-3}$ for electrons and at $n_\mathrm{B}>0.3$~fm$^{-3}$ for muons which is fairly enough for practical applications. Notice, that the approximation (\ref{eq:alpha_appr}) contains only the `direct' $R$-factors. This is because the contribution from the exchange diagrams is suppressed by weaker (logarithmic) propagator enhancement.

Very close to the direct Urca threshold, the difference $\varepsilon_N-\mu_N$ can become so small that the neutrino energy can not be neglected. In this region the treatment of the modified Urca rates becomes more involved.
The size of this special care region is  nevertheless rater small and can be neglected in the first approximation.

To finish the discussion, we illustrate the effect of the reconsidered modified Urca rates on the cooling of the non-superfluid neutron stars. The models of the NSs with non-superfluid nucleon cores which cool via the modified Urca process form the basis of the so-called standard cooling scenario of the NS cooling theory, e.g., \cite{YakovlevPethick2004ARA,Yakovlev2011MNRAS}. Weak density dependence of the modified Urca rates in the free OPE model of Ref.~\cite{FrimanMaxwell1979ApJ} leads to a consequent weak dependence of the cooling on the neutron star mass. Inclusion of the nuclear correlations does not change this conclusion appreciably. Results of the present work show that the modified Urca  emissivity is higher than in the model of Ref.~\cite{FrimanMaxwell1979ApJ} and in fact strongly depends on density, therefore one expects faster cooling with pronounced dependence on the NS mass. To perform simulations, we use the general relativistic cooling code described in Ref.~\cite{Gnedin2001MNRAS}. In the left panel of the Fig.~\ref{fig:alpha_e} we show the results of cooling simulations for the NS models with the equation of state consistent with our BHF calculations \cite{Sharma2015A&A}. For this EOS, the direct Urca process start to operate at $M_\mathrm{dU}\approx 1.3M_\odot$.
We plot cooling curves -- the dependencies of the surface temperature $T_s^\infty$ as seen by a distant observer on the NS age $t$ --  for NS models with masses from $1 M_\odot$ to $M_\mathrm{dU}$ stepped by $0.05M_\odot$. Group of black lines labelled `std' gives the old standard cooling results, while the group of red curves labelled `new' incorporates the rates calculated in the present work. Older curves are almost indistinguishable from each other, while the calculations in the present model show faster cooling and cooling curves fill broader region on the $T_s^\infty-t$ plane. Some cooling simulations, which were based on our preliminary results, can be also found in Ref.~\cite{Potekhin2018A&A}.

The direct Urca threshold for BHF EOS is relatively low, therefore the moderately massive non-superfluid NSs with such EOS cool fast. There are EOSs that predict higher direct Urca thresholds, so that NSs with broader range of masses cool down via the modified Urca process. Strictly speaking, it is not possible to use the rates calculated here with the different EOS, since the EOS and rate should be constructed self-consistently, based on the same nucleon interaction. However, the main effect pointed out here stems from the modification of the nucleon propagator that can be described in a universal way via the $R$-factors. As the results of the Fig.~\ref{fig:alpha_e} show, the effect of the modification of the nucleon interaction is smaller than the effect of the in-medium nucleon propagation. Based on that, we performed the model cooling calculations for another EOS using Eq.~(\ref{eq:alpha_appr}) for the rate. Specifically, we employed the model APR~I
from Ref.~\cite{Gusakov2005MNRAS} which is based on the analytical parameterization of the APR EOS \cite{Akmal1998PhRvC} proposed in Ref.~\cite{Heiselberg1999ApJ}. Direct Urca process for this EOS opens in relatively high-mass stars with $M_\mathrm{dU}=1.83 M_\odot$. The results of the cooling simulations are shown in the right panel in Fig.~\ref{fig:cool}. As expected, one observes much broader region filled by the cooling curves.

The broadening of the cooling curves region results from the strong increase in the modified Urca rates especially for the stars with masses lower, but close to $M_\mathrm{dU}$. However, Fig.~\ref{fig:cool} shows that the cooling is faster even for less massive stars, where central density is relatively far from the threshold density $n_\mathrm{dU}$. This decrease in the temperature can be roughly estimated from the approximate expression (\ref{eq:alpha_appr}) using the fact that the surface temperature scales as a  neutrino emissivity in power of $\approx -1/12$, see, e.g., Ref.~\cite{Yakovlev2011MNRAS}. Far from the threshold the density dependence of the modified Urca rate is modest (cf. Fig.~\ref{fig:alpha_e}) and  one can estimate the decrease of the temperature relative to the standard cooling scenario  as $\left(T_s^\infty\right)_{\mathrm{new}}/\left(T_s^\infty\right)_{
\mathrm{old}}\approx\left({\alpha_U^{\mathrm{approx}}}\right)^{-1/12}$, where $\alpha_U^{\mathrm{approx}}$ is taken at some typical intermediate density point (remember that $\alpha_U^{\mathrm{OPE}}\approx 1$). Therefore, typical values of $\alpha_U^{\mathrm{approx}}\sim 5-10$ (see Fig.~\ref{fig:alpha_e}) translate to 15--20\% decrease in the surface temperature of low-mass stars, which is indeed seen in Fig.~\ref{fig:cool}.

\section{Conclusions}
\label{S:conclusion}
We calculated the emissivity of the neutron branch of the modified Urca process in the non-superfluid neutron star cores based on the in-medium scattering matrix and taking into account the modification of the nucleon propagator in the beta-stable matter. We employed the non-relativistic quasiparticle approximation and $V-A$ model of the weak interaction. In-medium nucleon scattering was described in the BHF approach with account for the effective three-body forces. Our main conclusions are the following
\begin{itemize}
\item In the beta-stable matter, the modified Urca rates are strongly enhanced when the density approaches the direct Urca threshold density $n_\mathrm{dU}$. Sufficiently close to $n_\mathrm{dU}$, the modified Urca emissivity is inversely proportional to the distance from the dUrca threshold, $p_{\mathrm{F}n}-p_{\mathrm{F}p}-p_{\mathrm{F}\ell}$, measured in terms of momenta.
\item The use of the in-medium scattering matrix results in a certain reduction of the emissivities as compared to the free OPE model. However, in combination with the effect of the propagator amplification, the net result is by a factor of several higher than the free OPE one for all densities.
\item We constructed the simple practical expression (\ref{eq:alpha_appr}) that can be used in simulations for NS models employing any nuclear EOS.
\item NS cooling curves based on the updated rates differ significantly from the standard cooling curves. The cooling is faster, and shows more prominent dependence on the NS mass.
\end{itemize}

Our main conclusion is based solely on the beta-equilibrium condition and the quasi-particle approximation. We neglected the relativistic corrections, second-order processes, and the possible effects beyond the quasi-particle approximation. Nevertheless we think that this result survives more elaborated treatment, and it demonstrates that the inclusion of the potential energy of the nucleons in the medium mean field can have crucial impact on the resulting flavor changing reaction rates. Notice, that   in the context of the charged current opacity in the hot supernova matter, the potential energy difference between neutron and proton quasipartcles was shown to be important for the charged current neutrino emissivities and opacities from the dUrca processes in dense and hot
supernova matter (see Ref. \cite{RobertsReddy2017PhRvC}, where previous work is critically reviewed).

Here we discussed only the neutron branch of the modified Urca process. The similar modifications are expected for the weaker proton branch [$N=p$ in (\ref{eq:murca})]. In a first approximation, one can use a simple scaling relations (see eq.~(142) in Ref.~\cite{Yakovlev2001physrep}) to find p-branch mUrca emissivity from the n-branch emissivity. The detailed consideration of the proton branch, as well as of the effects of different nuclear potentials and/or the models of three-body forces is left for the future studies.

Our results can be important not only for the cooling of the isolated neutron stars, but also for the thermal states of NSs in low-mass X-ray binaries (soft X-ray transients) \cite{YakovlevPethick2004ARA,Fortin2018MNRAS}. Flavor-changing reactions also  contribute to the bulk viscosity of the matter (e.g., \cite{Schmitt17}) that can be important in the evolution of the NS oscillations. Finally, these reactions are important in regulating the magnetic field evolution in the NSs where some degree of the compositional asymmetry induced by the magnetic field can be expected, e.g., \cite{Gusakov2017PhRvD}.

Finally, here we neglected the effects of the nucleon superfluidity. There is currently a little doubt that the nucleons in the NS cores can be in the paired state. However, the critical temperatures calculated in various approaches are strongly density dependent and these dependencies are highly uncertain. At temperature sufficiently lower than either the neutron (triplet) or the proton (singlet) critical temperature, both direct and modified Urca rates are suppressed exponentially due to the gap-restricted phasespace, and the processes which involve ungapped species dominate. Only in the region of the star where the superfluidity is not well-developed, the modified Urca process is important and the enhancement found here needs to be taken into account. In this case, however, the situation becomes more complicated since the single particle spectrum of superfluid species contains the gap which shifts the position of the direct Urca pole from the Fermi surface. Fortunately, in the discussed regions, the gap value(s) are of the order of temperature and as such they can be neglected in the first approximation in the quasiparticls propagators and in the scattering matrices so that Eq.~(\ref{eq:alpha_appr}) can still be used supplied with the superfluid suppression factors \cite{Yakovlev2001physrep}. Notice, that in the same region, the situation is further complicated by the appearance of the neutrino emission associated with the Cooper pair formation processes \cite{Page2004ApJS,Gusakov2004A&A,Shternin2015MNRAS,Schmitt17}.
The detailed treatment of the superfluid case is  outside the scope of the present paper and we plan to address the effects of superfluidity in the future work.

\section*{Acknowledgments}
The work of PS was supported by RFBR, grant No. 16-32-00507~mol$\_$a and the Foundation for the Advancement of Theoretical Physics and Mathematics ``BASIS''. The part of this work was completed during the STSM mission
ECOST-STSM-MP1304-061014-049724 of the COST NewCompstar project.  This
work was partially supported by the Polish NCN research grant  OPUS7  no. 2014/13/B/ST9/02621.
We thank A.~D. Kaminker, E.~E. Kolomeitsev, S. Reddy, and A.~D.
Sedrakian for valuable discussions.

\def\apj{Astroph. J.}
\def\apjl{Astroph. J. Lett.}
\def\apjs{Astroph. J. Suppl.}
\def\mnras{Mon. Not. Roy. Astron. Soc.}
\def\nat{Nature}
\def\aap{Astron. Astroph.}
\def\pr{Phys. Rev.}
\def\prc{Phys. Rev. {\rm C}}
\def\prd{Phys. Rev. {\rm D}}
\def\plb{Physics Letters {\rm B}}
\def\apss{Astroph. Space Sci.}
\def\pla{Physics Letters {\rm A}}
\def\ssr{Space Sci. Rev.}
\def\araa{Ann. Rev. Astron. Astroph.}
\def\aj{Astron. J.}
\def\jphys{J. Phys.}
\def\npa{Nuclear Physics {\rm A}}
\def\npb{Nuclear Physics {\rm B}}
\def\ijmpe{International Journal of Modern Physics {\rm E}}
\def\ijmpd{International Journal of Modern Physics {\rm D}}
\def\ijmpa{International Journal of Modern Physics {\rm A}}
\def\physrep{Physics Reports}



\begin{thebibliography}{10}
\expandafter\ifx\csname url\endcsname\relax
  \def\url#1{\texttt{#1}}\fi
\expandafter\ifx\csname urlprefix\endcsname\relax\def\urlprefix{URL }\fi
\expandafter\ifx\csname href\endcsname\relax
  \def\href#1#2{#2} \def\path#1{#1}\fi

\bibitem{HPY2007Book}
P.~{Haensel}, A.~Y. {Potekhin}, D.~G. {Yakovlev}, {Neutron Stars 1: Equation of
  State and Structure}, Vol. 326 of Astrophysics and Space Science Library,
  Springer Science+Buisness Media, New York, 2007.

\bibitem{Degenaar2018arXiv}
N.~{Degenaar}, V.~F. {Suleimanov}, {Testing the equation of state of neutron
  stars with electromagnetic observations}, ArXiv e-prints (2018) 1806.02833.

\bibitem{Abbott2017ApJ}
B.~P. {Abbott}, R.~{Abbott}, T.~D. {Abbott}, F.~{Acernese}, K.~{Ackley},
  C.~{Adams}, T.~{Adams}, P.~{Addesso}, R.~X. {Adhikari}, V.~B. {Adya}, et~al.,
  {Multi-messenger Observations of a Binary Neutron Star Merger}, \apjl 848
  (2017) L12.

\bibitem{Abbott2017PhRvL}
B.~P. {Abbott}, R.~{Abbott}, T.~D. {Abbott}, F.~{Acernese}, K.~{Ackley},
  C.~{Adams}, T.~{Adams}, P.~{Addesso}, R.~X. {Adhikari}, V.~B. {Adya}, et~al.,
  {GW170817: Observation of Gravitational Waves from a Binary Neutron Star
  Inspiral}, Physical Review Letters 119~(16) (2017) 161101.

\bibitem{YakovlevPethick2004ARA}
D.~G. {Yakovlev}, C.~J. {Pethick}, {Neutron Star Cooling}, Ann. Rev. Astron.
  Astrophys. 42 (2004) 169--210.

\bibitem{Yakovlev2001physrep}
D.~G. {Yakovlev}, A.~D. {Kaminker}, O.~Y. {Gnedin}, P.~{Haensel}, {Neutrino
  emission from neutron stars}, Phys.~Rep. 354 (2001) 1--155.

\bibitem{FrimanMaxwell1979ApJ}
B.~L. {Friman}, O.~V. {Maxwell}, {Neutrino emissivities of neutron stars}, \apj
  232 (1979) 541--557.

\bibitem{Schmitt17}
A.~{Schmitt}, P.~{Shternin}, {Reaction rates and transport in neutron stars},
  ArXiv e-prints (2017) 1711.06520.

\bibitem{Blaschke1995MNRAS}
D.~{Blaschke}, G.~{Ropke}, H.~{Schulz}, A.~D. {Sedrakian}, D.~N.
  {Voskresensky}, {Nuclear in-medium effects and neutrino emissivity of neutron
  stars}, \mnras 273 (1995) 596--602.

\bibitem{Niri2016PhRvC}
A.~{Dehghan Niri}, H.~R. {Moshfegh}, P.~{Haensel}, {Nuclear correlations and
  neutrino emissivity from the neutron branch of the modified Urca process},
  \prc 93~(4) (2016) 045806.

\bibitem{Haensel1987A&A}
P.~{Haensel}, A.~J. {Jerzak}, {Mean free paths of non-degenerate neutrinos in
  neutron star matter}, \aap 179 (1987) 127--133.

\bibitem{Voskresensky1986JETP}
D.~N. {Voskresensky}, A.~V. {Senatorov}, {Neutrino emission by neutron stars},
  J. Exp. Theor. Phys. 63 (1986) 885.

\bibitem{Voskresensky2001LNP}
D.~N. {Voskresensky}, {Neutrino Cooling of Neutron Stars: Medium Effects}, in:
  D.~{Blaschke}, N.~K. {Glendenning}, A.~{Sedrakian} (Eds.), Physics of Neutron
  Star Interiors, Vol. 578 of Lecture Notes in Physics, Berlin Springer Verlag,
  2001, p. 467.

\bibitem{Migdal1990PhR}
A.~B. {Migdal}, E.~E. {Saperstein}, M.~A. {Troitsky}, D.~N. {Voskresensky},
  {Pion degrees of freedom in nuclear matter}, Phys. Rep. 192 (1990) 179--437.

\bibitem{Kaminker2016Ap&SS}
A.~D. {Kaminker}, D.~G. {Yakovlev}, P.~{Haensel}, {Theory of neutrino emission
  from nucleon-hyperon matter in neutron stars: angular integrals}, \apss 361
  (2016) 267.

\bibitem{Timmermans2002PhRvC}
R.~G. {Timmermans}, A.~Y. {Korchin}, E.~N. {van Dalen}, A.~E. {Dieperink},
  {Soft electroweak bremsstrahlung: Theorems and astrophysical relevance}, \prc
  65~(6) (2002) 064007.

\bibitem{Hanhart2001PhLB}
C.~{Hanhart}, D.~R. {Phillips}, S.~{Reddy}, {Neutrino and axion emissivities of
  neutron stars from nucleon-nucleon scattering data}, \plb 499 (2001) 9--15.

\bibitem{Baldo1999Book}
M.~Baldo (Ed.), Nuclear Methods and the Nuclear Equation of State, Vol.~8 of
  International Review of Nuclear Physics, World Scientific, Singapore, 1999.

\bibitem{Wiringa1995PhRvC}
R.~B. {Wiringa}, V.~G.~J. {Stoks}, R.~{Schiavilla}, {Accurate nucleon-nucleon
  potential with charge-independence breaking}, \prc 51 (1995) 38--51.

\bibitem{Grange1989}
P.~{Grang{\'e}}, A.~{Lejeune}, M.~{Martzolff}, J.-F. {Mathiot}, {Consistent
  three-nucleon forces in the nuclear many-body problem}, \prc 40 (1989)
  1040--1060.

\bibitem{Carlson1983NuPhA}
J.~{Carlson}, V.~R. {Pandharipande}, R.~B. {Wiringa}, {Three-nucleon
  interaction in 3-, 4- and {$\infty$}-body systems}, Nuclear Physics A 401
  (1983) 59--85.

\bibitem{Baldo2008PhLB}
M.~{Baldo}, A.~E. {Shaban}, {Dependence of the nuclear equation of state on
  two-body and three-body forces}, Physics Letters B 661 (2008) 373--377.

\bibitem{Baldo2014PhRvC}
M.~{Baldo}, G.~F. {Burgio}, H.-J. {Schulze}, G.~{Taranto}, {Nucleon effective
  masses within the Brueckner-Hartree-Fock theory: Impact on stellar neutrino
  emission}, \prc 89~(4) (2014) 048801.

\bibitem{Sharma2015A&A}
B.~K. {Sharma}, M.~{Centelles}, X.~{Vi{\~n}as}, M.~{Baldo}, G.~F. {Burgio},
  {Unified equation of state for neutron stars on a microscopic basis}, \aap
  584 (2015) A103.

\bibitem{Sedrakian2007PrPNP}
A.~{Sedrakian}, {The physics of dense hadronic matter and compact stars},
  Progress in Particle and Nuclear Physics 58 (2007) 168--246.
\newblock \href {http://dx.doi.org/10.1016/j.ppnp.2006.02.002}
  {\path{doi:10.1016/j.ppnp.2006.02.002}}.

\bibitem{VanDalen2003PhRvC}
E.~N. {van Dalen}, A.~E. {Dieperink}, J.~A. {Tjon}, {Neutrino emission in
  neutron stars}, \prc 67~(6) (2003) 065807.

\bibitem{BrownRho2004}
G.~E. {Brown}, M.~{Rho}, {Matching the QCD and hadron sectors and
  medium-dependent meson masses; hadronization in relativistic heavy ion
  collisions}, \physrep 398 (2004) 301--325.
\newblock \href {http://arxiv.org/abs/nucl-th/0206021}
  {\path{arXiv:nucl-th/0206021}}, \href
  {http://dx.doi.org/10.1016/j.physrep.2004.05.006}
  {\path{doi:10.1016/j.physrep.2004.05.006}}.

\bibitem{Dong2016ApJ}
J.~M. {Dong}, U.~{Lombardo}, H.~F. {Zhang}, W.~{Zuo}, {Role of Nucleonic Fermi
  Surface Depletion in Neutron Star Cooling}, \apj 817 (2016) 6.

\bibitem{Schwenk2003NuPhA}
A.~{Schwenk}, B.~{Friman}, G.~E. {Brown}, {Renormalization group approach to
  neutron matter: quasiparticle interactions, superfluid gaps and the equation
  of state}, \npa 713 (2003) 191--216.

\bibitem{Schwenk2004PhLB}
A.~{Schwenk}, P.~{Jaikumar}, C.~{Gale}, {Neutrino bremsstrahlung in neutron
  matter from effective nuclear interactions}, \plb 584 (2004) 241--250.

\bibitem{Yakovlev2011MNRAS}
D.~G. {Yakovlev}, W.~C.~G. {Ho}, P.~S. {Shternin}, C.~O. {Heinke}, A.~Y.
  {Potekhin}, {Cooling rates of neutron stars and the young neutron star in the
  Cassiopeia A supernova remnant}, \mnras 411 (2011) 1977--1988.

\bibitem{Gnedin2001MNRAS}
O.~Y. {Gnedin}, D.~G. {Yakovlev}, A.~Y. {Potekhin}, {Thermal relaxation in
  young neutron stars}, \mnras 324 (2001) 725--736.

\bibitem{Potekhin2018A&A}
A.~Y. {Potekhin}, G.~{Chabrier}, {Magnetic neutron star cooling and
  microphysics}, \aap 609 (2018) A74.

\bibitem{Gusakov2005MNRAS}
M.~E. {Gusakov}, A.~D. {Kaminker}, D.~G. {Yakovlev}, O.~Y. {Gnedin}, {The
  cooling of Akmal-Pandharipande-Ravenhall neutron star models}, \mnras 363
  (2005) 555--562.

\bibitem{Akmal1998PhRvC}
A.~{Akmal}, V.~R. {Pandharipande}, D.~G. {Ravenhall}, {Equation of state of
  nucleon matter and neutron star structure}, \prc 58 (1998) 1804--1828.

\bibitem{Heiselberg1999ApJ}
H.~{Heiselberg}, M.~{Hjorth-Jensen}, {Phase Transitions in Neutron Stars and
  Maximum Masses}, \apjl 525 (1999) L45--L48.

\bibitem{RobertsReddy2017PhRvC}
L.~F. {Roberts}, S.~{Reddy}, {Charged current neutrino interactions in hot and
  dense matter}, \prc 95~(4) (2017) 045807.

\bibitem{Fortin2018MNRAS}
M.~{Fortin}, G.~{Taranto}, G.~F. {Burgio}, P.~{Haensel}, H.-J. {Schulze}, J.~L.
  {Zdunik}, {Thermal states of neutron stars with a consistent model of
  interior}, \mnras 475 (2018) 5010--5022.

\bibitem{Gusakov2017PhRvD}
M.~E. {Gusakov}, E.~M. {Kantor}, D.~D. {Ofengeim}, {Evolution of the magnetic
  field in neutron stars}, \prd 96~(10) (2017) 103012.

\bibitem{Page2004ApJS}
D.~{Page}, J.~M. {Lattimer}, M.~{Prakash}, A.~W. {Steiner}, {Minimal Cooling of
  Neutron Stars: A New Paradigm}, \apjs 155 (2004) 623--650.

\bibitem{Gusakov2004A&A}
M.~E. {Gusakov}, A.~D. {Kaminker}, D.~G. {Yakovlev}, O.~Y. {Gnedin}, {Enhanced
  cooling of neutron stars via Cooper-pairing neutrino emission}, \aap 423
  (2004) 1063--1071.

\bibitem{Shternin2015MNRAS}
P.~S. {Shternin}, D.~G. {Yakovlev}, {Self-similarity relations for cooling
  superfluid neutron stars}, \mnras 446 (2015) 3621--3630.

\end{thebibliography}

\end{document}